\documentclass[aps,prb,twocolumn,amsmath,amssymb,groupedaddress,longbibliography
]{revtex4-2}

\usepackage{graphicx}
\usepackage{dcolumn}
\usepackage{bm}
\usepackage{multirow}
\usepackage{braket}
\usepackage{longtable}
\usepackage{color}
\usepackage{ulem}
\usepackage{booktabs}

\begin{document}

\title{
Configuration determination for chiral and polar crystals\\
by anisotropic NMR shift
}

\author{Hiroaki Kusunose$^{1,2}$ and Jun Kikuchi$^{1}$}
\affiliation{
$^1$Department of Physics, Meiji University, Kawasaki 214-8571, Japan \\
$^2$Quantum Research Center for Chirality, Institute for Molecular Science, Okazaki 444-8585, Japan
}

\begin{abstract}
We propose a method to perform a configuration determination for chiral and/or polar crystals by utilizing anisotropy of NMR shift.
The chirality (handedness) or polarity of a crystal, that is characterized by its sign, can be extracted from the asymmetric magnetic-field angle dependence of NMR shift in the appropriate plane, as the configuration is reflected in the off-diagonal components of the shift tensor.
This method is applicable to the triclinic, monoclinic, and trigonal crystal structures among 230 crystallographic space groups, and the appropriate planes to extract the asymmetric field-angle dependence are tabulated for all relevant space groups.
We discuss how to determine the appropriate plane, and to identify each contribution of twin domains, and so on, by using the specific examples of Te, IrSn$_{4}$ and RhSn$_{4}$, and the spontaneous symmetry-breaking phases of URhSn.
We also argue that an \textit{absolute}-configuration determination is also possible in accordance with the relation between chirality or polarity of crystal and the sign of the relevant off-diagonal component of the shift tensor, provided by theoretical evaluation of internal magnetic field from surrounding ions, and/or by experimental input on the shape of the asymmetric etch pit on the cleaved face of crystal.
\end{abstract}

\maketitle

\section{Introduction}

Inversion-symmetry breaking together with spin-orbit coupling has been attracted much attention as they are fundamental source to bring about various nonreciprocal phenomena and cross-correlated responses~\cite{Nagaosa23,Togawa23,Yu23}.
Among them, polar crystals provide typical playgrounds, in which some physical properties of materials, including advanced applications, such as pyroelectricity, piezoelectricity, ferroelectricity, second harmonic generation, and electrooptic effect, are only allowed, or they are strongly enhanced.
Moreover, chiral crystals add an ability to convert from axial quantities to polar ones and vice versa, such as a conversion from the applied electric current to the angular momentum~\cite{Yoda15,Yoda18,Shalygin12,Furukawa17,Furukawa21,Oiwa22,Suzuki23}.
It is remarkable that the spin degree of freedom is controlled even in nonmagnetic chiral materials, which is the so-called ``Chirality-Induced Spin Selectivity'' (CISS)~\cite{Goehler11,Naaman12,Naaman19,Naaman20,Evers22,Inui20,Nabei20,Shiota21}.

Polarity of a crystal is determined by the sign of the polarization.
Similarly, chirality (handedness) of a crystal is determined by the sign of the ``chiralization'', which is recently introduced quantitatively by using the concept of the electric-toroidal monopole that corresponds to a sort of ``order parameter'' of the chirality~\cite{Kelvin04,Wagniere07,Barron04,Oiwa22,Kishine22,Hayami18,Kusunose20}.
Since the nonreciprocal phenomena and cross-correlated responses in polar or chiral crystals depend essentially on their polarity or chirality, it is fundamentally significant to make a configuration determination of samples.

Circular dichroism is a very efficient method to determine the configuration of chirality, although there are several difficulties~\cite{Perez22}.
Single crystal X-ray diffraction is also useful, and the Flack parameter is a standard indicator of chirality or polarity~\cite{Flack83,Flack00,Flack08}.
Nevertheless, these methods sometime do not work out because of sample quality, size, and resolution of X-ray beam, and so on.
Therefore, a complementary method of a configuration determination for chirality or polarity is required.

In this paper, we propose a method to perform a configuration determination for chiral and/or polar crystals by using the anisotropic magnetic-field angle dependence of NMR shift.
The chirality (handedness) or polarity of a crystal, that is characterized by its sign, can be extracted from the asymmetric field-angle dependence in the appropriate plane, as the configuration is reflected in the off-diagonal components of the shift tensor at the nuclear site.

This paper is organized as follows.
In Sec.~II, we present the fundamental expressions of the NMR shift, and decompose the components of the shift and electric-field gradient (EFG) tensors by using the orthogonal basis set.
In the decomposed expression, the symmetry of the shift and EFG tensors become apparent which are tabulated for all site symmetries of nuclear sites.
In Sec.~III, we discuss how to determine the appropriate plane from group theoretical point of view, and summarize all appropriate planes of the relevant space groups (triclinic, monoclinic, and trigonal crystal systems) to perform a configuration determination.
In Sec.~IV, we demonstrate the present method by using the specific examples of Te (high-symmetry Wyckoff position in trigonal system), IrSn$_{4}$ and RhSn$_{4}$ (general Wyckoff position in trigonal system), and the spontaneous symmetry-breaking phases of URhSn (high-symmetry Wyckoff position in ordered chiral or polar system).
Section~V summarizes the paper.
We have two appendices.
The expressions of shift and EFG tensors and their first-order perturbation forms are given in Appendix A.
The algorithm for fitting of experimental data is briefly explained in Appendix B.

\section{Anisotropy of NMR shift}

\subsection{Nuclear spin Hamiltonian and NMR shift}

Let us begin with the nuclear spin Hamiltonian to discuss the anisotropy of the NMR shift~\cite{Abragam61}.
The Hamiltonian for the nuclear spin $\bm{I}_{j}$ (dimensionless) at $j$ site is expressed as
\begin{align}
&
H^{(j)}=\hbar\omega_{0}\sum_{\alpha\beta}\left[-h^{\alpha}(\delta_{\alpha\beta}+K_{\alpha\beta}^{(j)})I_{j}^{\beta}+\widetilde{V}_{\alpha\beta}^{(j)}Q_{j}^{\alpha\beta}\right],
\cr&\quad
Q_{j}^{\alpha\beta}=\frac{1}{2}(I_{j}^{\alpha}I_{j}^{\beta}+I_{j}^{\beta}I_{j}^{\alpha})-\frac{1}{3}I(I+1)\delta_{\alpha\beta},
\cr&\quad
\widetilde{V}_{\alpha\beta}^{(j)}=\frac{eQ/\hbar\omega_{0}}{2I(2I-1)}V_{\alpha\beta}^{(j)},
\label{eq:ham}
\end{align}
where the first, second, and third terms are the Zeeman coupling, hyperfine interaction, and quadrupolar interaction with the quadrupole moment $Q$, respectively.
The second-rank operator $Q_{j}^{\alpha\beta}$ is the electric quadrupole operator, which is finite only when the nuclear spin $I>1/2$.
Here, $\bm{h}=\bm{H}/|\bm{H}|$ is the unit vector along the applied magnetic-field direction, and $\omega_{0}=\gamma_{\rm n}|\bm{H}|$ is the nuclear resonance frequency with the nuclear gyromagnetic ratio $\gamma_{\rm n}$.
The coefficients $K_{\alpha\beta}^{(j)}$ and $V_{\alpha\beta}^{(j)}$ are the shift and EFG tensors, respectively, which are both time-reversal even second-rank polar tensor with real components, and the latter is traceless symmetric.

The eigenvalues of the Hamiltonian (\ref{eq:ham}) are obtained by diagonalizing $(2I+1)\times(2I+1)$ matrix for the basis ($m=I,I-1,\cdots,-I$), and we denote them as $\hbar\Omega_{m}^{(j)}$ in ascending order.
The label $m$ of the eigenvalues is assigned so that they are reduced to those of the Zeeman coupling in the noninteracting limit.
The total shift of the resonance frequency relative to the bare one in the presence of the hyperfine and quadrupolar interactions is defined by
\begin{align}
T_{m}^{(j)}\equiv\frac{\Omega_{m-1}^{(j)}-\Omega_{m}^{(j)}-\omega_{0}}{\omega_{0}}
\quad
(m=I,I-1,\cdots,-I+1).
\label{eq:shift}
\end{align}
The contribution of the hyperfine interaction to $T_{m}^{(j)}$ is often called the Knight shift.

When the hyperfine and quadrupolar interactions are weak enough as compared with the Zeeman coupling, we evaluate the eigenvalues by the first-order perturbation as
\begin{align}
\frac{\Omega_{m}^{(j)}}{\omega_{0}}=-m-\sum_{\alpha\beta}\left(K_{\alpha\beta}^{(j)}m-\frac{\widetilde{V}_{\alpha\beta}^{(j)}}{2}[3m^{2}-I(I+1)]\right)h^{\alpha}h^{\beta}.
\end{align}
Using this expression, the total shift (\ref{eq:shift}) is given by
\begin{align}
T_{m}^{(j)}=\sum_{\alpha\beta}\left[\frac{1}{2}(K_{\alpha\beta}^{(j)}+K_{\beta\alpha}^{(j)})-3\widetilde{V}_{\alpha\beta}^{(j)}\left(m-\frac{1}{2}\right)\right]h^{\alpha}h^{\beta}.
\label{eq:shiftpb}
\end{align}
Note that in the first-order perturbation, only the symmetric components of $K_{\alpha\beta}^{(j)}$ contribute to the total shift as $h^{\alpha}h^{\beta}$ is symmetric with respect to $\alpha\leftrightarrow\beta$.
Then, the expansion coefficients, $(k_{x}',k_{y}',k_{z}')$, as defined below can be ignored.

From this expression, it is evident that an asymmetric field-angle dependence appears in the presence of the off-diagonal components.
It is the key ingredient to perform a configuration determination for chirality and/or polarity of crystal as discussed later.

\subsection{Decomposition of shift and EFG tensors}

In general, $3\times 3$ time-reversal even second-rank polar tensor can be decomposed as the linear combination of nine bases composed of the electric monopole ($\mathbb{Q}_{0}$), electric quadupoles ($\mathbb{Q}_{U}$, $\mathbb{Q}_{V}$, $\mathbb{Q}_{YZ}$, $\mathbb{Q}_{ZX}$, $\mathbb{Q}_{XY}$), and the electric-toroidal dipoles ($\mathbb{G}_{X}$, $\mathbb{G}_{Y}$, $\mathbb{G}_{Z}$), and they are defined as follows ($\mathbb{G}_{\alpha}=i\mathbb{M}_{\alpha}$),
\begin{align}
&
\mathbb{Q}_{0}=\frac{1}{\sqrt{3}}\begin{pmatrix}
1 & 0 & 0 \\
0 & 1 & 0 \\
0 & 0 & 1
\end{pmatrix},
\quad
\mathbb{Q}_{U}=\frac{1}{\sqrt{6}}\begin{pmatrix}
-1 & 0 & 0 \\
0 & -1 & 0 \\
0 & 0 & 2
\end{pmatrix},
\cr&
\mathbb{Q}_{V}=\frac{1}{\sqrt{2}}\begin{pmatrix}
1 & 0 & 0 \\
0 & -1 & 0 \\
0 & 0 & 0
\end{pmatrix}
\quad
\mathbb{Q}_{YZ}=\frac{1}{\sqrt{2}}\begin{pmatrix}
0 & 0 & 0 \\
0 & 0 & 1 \\
0 & 1 & 0
\end{pmatrix},
\cr&
\mathbb{Q}_{ZX}=\frac{1}{\sqrt{2}}\begin{pmatrix}
0 & 0 & 1 \\
0 & 0 & 0 \\
1 & 0 & 0
\end{pmatrix},
\quad
\mathbb{Q}_{XY}=\frac{1}{\sqrt{2}}\begin{pmatrix}
0 & 1 & 0 \\
1 & 0 & 0 \\
0 & 0 & 0
\end{pmatrix},
\cr&
\mathbb{M}_{X}=\frac{1}{\sqrt{2}}\begin{pmatrix}
0 & 0 & 0 \\
0 & 0 & -i \\
0 & i & 0
\end{pmatrix},
\quad
\mathbb{M}_{Y}=\frac{1}{\sqrt{2}}\begin{pmatrix}
0 & 0 & i \\
0 & 0 & 0 \\
-i & 0 & 0
\end{pmatrix},
\cr&
\mathbb{M}_{Z}=\frac{1}{\sqrt{2}}\begin{pmatrix}
0 & -i & 0 \\
i & 0 & 0 \\
0 & 0 & 0
\end{pmatrix},
\end{align}
where $U=3Z^{2}-R^{2}$ and $V=X^{2}-Y^{2}$.
They are hermitian matrices, and satisfy the orthonormality as ${\rm Tr}(\mathbb{X}_{i}\mathbb{X}_{j})=\delta_{ij}$ ($\mathbb{X}=\mathbb{Q},\mathbb{M}$).
Here we have used the capital letters $(XYZ)$ representing a local coordinate at a nuclear site.

By using these bases, the shift tensor is decomposed as
\begin{align}
\hat{K}&=
\sum_{i}^{0,U,V,YZ,ZX,XY}k_{i}\,\mathbb{Q}_{i}+\sum_{i}^{X,Y,Z}k_{i}'\,\mathbb{G}_{i}
\cr&
=\begin{pmatrix}
\frac{k_{0}}{\sqrt{3}}-\frac{k_{U}}{\sqrt{6}}+\frac{k_{V}}{\sqrt{2}} & \frac{k_{Z}+k_{Z}'}{\sqrt{2}} & \frac{k_{Y}-k_{Y}'}{\sqrt{2}} \\
\frac{k_{Z}-k_{Z}'}{\sqrt{2}} & \frac{k_{0}}{\sqrt{3}}-\frac{k_{U}}{\sqrt{6}}-\frac{k_{V}}{\sqrt{2}} & \frac{k_{X}+k_{X}'}{\sqrt{2}} \\
\frac{k_{Y}+k_{Y}'}{\sqrt{2}} & \frac{k_{X}-k_{X}'}{\sqrt{2}} & \frac{k_{0}}{\sqrt{3}}+\frac{2k_{U}}{\sqrt{6}}
\end{pmatrix},
\quad
\end{align}
where the coefficients are denoted as $(k_{YZ},k_{ZX},k_{XY})\to (k_{X},k_{Y},k_{Z})$ for notational simplicity, and they can be obtained by $k_{i}={\rm Tr}(\hat{K}\mathbb{Q}_{i})$ and $k_{i}'=-{\rm Tr}(\hat{K}\mathbb{G}_{i})$ as
\begin{align}
&
k_{0}=\frac{1}{\sqrt{3}}(K_{XX}+K_{YY}+K_{ZZ}),
\cr&
k_{U}=\frac{1}{\sqrt{6}}(2K_{ZZ}-K_{XX}-K_{YY}),
\quad
k_{V}=\frac{1}{\sqrt{2}}(K_{XX}-K_{YY}),
\cr&
k_{X}=\frac{1}{\sqrt{2}}(K_{YZ}+K_{ZY}),
\quad
k_{Y}=\frac{1}{\sqrt{2}}(K_{ZX}+K_{XZ}),
\cr&
k_{Z}=\frac{1}{\sqrt{2}}(K_{XY}+K_{YX}),
\quad
k_{X}'=\frac{1}{\sqrt{2}}(K_{YZ}-K_{ZY}),
\cr&
k_{Y}'=\frac{1}{\sqrt{2}}(K_{ZX}-K_{XZ}),
\quad
k_{Z}'=\frac{1}{\sqrt{2}}(K_{XY}-K_{YX}).
\end{align}

\begin{table*}
\caption{Site symmetry and active component of the multipole bases. The primary axis of the local coordinate is taken as $Z$ axis.
The abbreviations, $U=2Z^{2}-X^{2}-Y^{2}$ and $V=X^{2}-Y^{2}$, are used.}
\label{tbl:activeq}
\centering{
\begin{tabular}{c|c|c|c|cccccc|ccc} \hline\hline
Crystal system & Site & International & \# acitve & $\mathbb{Q}_0$ & $\mathbb{Q}_U$ & $\mathbb{Q}_V$ & $\mathbb{Q}_{YZ}$ & $\mathbb{Q}_{ZX}$ & $\mathbb{Q}_{XY}$ & $\mathbb{G}_{X}$ & $\mathbb{G}_{Y}$ & $\mathbb{G}_{Z}$ \\
& symmetry & representation & symmetric+ & $(k_0)$ & $(k_U)$ & $(k_V)$ & $(k_{X})$ & $(k_{Y})$ & $(k_{Z})$ & $(k_{X}')$ & $(k_{Y}')$ & $(k_{Z}')$ \\
& & & anti-symmetric & & $(v_U)$ & $(v_V)$ & $(v_{X})$ & $(v_{Y})$ & $(v_{Z})$ & & & \\ \hline
triclinic & C$_{\rm 1}$ & $1$ & $6+3$ & $\checkmark$ & $\checkmark$ & $\checkmark$ & $\checkmark$ & $\checkmark$ & $\checkmark$ & $\checkmark$ & $\checkmark$ & $\checkmark$ \\
& C$_{\rm i}$ & $\bar{1}$ & $6+3$ &  $\checkmark$ & $\checkmark$ & $\checkmark$ & $\checkmark$ & $\checkmark$ & $\checkmark$ & $\checkmark$ & $\checkmark$ & $\checkmark$ \\ \hline
monoclinic & C$_{\rm 2}$ & $2$ & $4+1$ & $\checkmark$ & $\checkmark$ & $\checkmark$ & & & $\checkmark$ & & & $\checkmark$ \\
& C$_{\rm s}$ & $m$ & $4+1$ & $\checkmark$ & $\checkmark$ & $\checkmark$ & & & $\checkmark$ & & & $\checkmark$ \\
& C$_{\rm 2h}$ & $2/m$ & $4+1$ & $\checkmark$ & $\checkmark$ & $\checkmark$ & & & $\checkmark$ & & & $\checkmark$ \\ \hline
orthorhombic & D$_{\rm 2}$ & $222$ & 3 & $\checkmark$ & $\checkmark$ & $\checkmark$ & & & & & & \\
& C$_{\rm 2v}$ & mm2 & 3 & $\checkmark$ & $\checkmark$ & $\checkmark$ & & & & & & \\
& D$_{\rm 2h}$ & mmm & 3 & $\checkmark$ & $\checkmark$ & $\checkmark$ & & & & & & \\ \hline
tetragonal & C$_{\rm 4}$ & $4$ & $2+1$ & $\checkmark$ & $\checkmark$ & & & & & & & $\checkmark$ \\
& S$_{\rm 4}$ & $\bar{4}$ & $2+1$ & $\checkmark$ & $\checkmark$ & & & & & & & $\checkmark$ \\
& C$_{\rm 4h}$ & 4/m & $2+1$ & $\checkmark$ & $\checkmark$ & & & & & & & $\checkmark$ \\
& D$_{\rm 4}$ & 422 & 2 & $\checkmark$ & $\checkmark$ & & & & & & & \\
& C$_{\rm 4v}$ & 4mm & 2 & $\checkmark$ & $\checkmark$ & & & & & & & \\
& D$_{\rm 2d}$ & $\bar{4}$2m & 2 & $\checkmark$ & $\checkmark$ & & & & & & & \\
& D$_{\rm 4h}$ & 4/mmm & 2 & $\checkmark$ & $\checkmark$ & & & & & & & \\ \hline
trigonal & C$_{\rm 3}$ & $3$ & $2+1$ & $\checkmark$ & $\checkmark$ & & & & & & & $\checkmark$ \\
& C$_{\rm 3i}$ & $\bar{3}$ & $2+1$ & $\checkmark$ & $\checkmark$ & & & & & & & $\checkmark$ \\
& D$_{\rm 3}$ & 32 & 2 & $\checkmark$ & $\checkmark$ & & & & & & & \\
& C$_{\rm 3v}$ & 3m & 2 & $\checkmark$ & $\checkmark$ & & & & & & & \\
& D$_{\rm 3d}$ & $\bar{3}$m & 2 & $\checkmark$ & $\checkmark$ & & & & & & & \\ \hline
hexagonal & C$_{\rm 6}$ & 6 & $2+1$ & $\checkmark$ & $\checkmark$ & & & & & & & $\checkmark$ \\
& C$_{\rm 3h}$ & $\bar{6}$ & $2+1$ & $\checkmark$ & $\checkmark$ & & & & & & & $\checkmark$ \\
& C$_{\rm 6h}$ & 6/m & $2+1$ & $\checkmark$ & $\checkmark$ & & & & & & & $\checkmark$ \\
& D$_{\rm 6}$ & 622 & 2 & $\checkmark$ & $\checkmark$ & & & & & & & \\
& C$_{\rm 6v}$ & 6mm & 2 & $\checkmark$ & $\checkmark$ & & & & & & & \\
& D$_{\rm 3h}$ & $\bar{6}$2m & 2 & $\checkmark$ & $\checkmark$ & & & & & & & \\
& D$_{\rm 6h}$ & 6/mmm & 2 & $\checkmark$ & $\checkmark$ & & & & & & & \\ \hline
cubic & T & 23 & 1 & $\checkmark$ & & & & & & & & \\
& T$_{\rm h}$ & m$\bar{3}$ & 1 & $\checkmark$ & & & & & & & & \\
& O & 432 & 1 & $\checkmark$ & & & & & & & & \\
& T$_{\rm d}$ & $\bar{4}$3m & 1 & $\checkmark$ & & & & & & & & \\
& O$_{\rm h}$ & m$\bar{3}$m & 1 & $\checkmark$ & & & & & & & & \\
\hline\hline
\end{tabular}
}
\end{table*}

The EFG tensor $\widetilde{V}_{\alpha\beta}$ can also be decomposed in a similar way, and we denote the coefficients as $v_{i}$.
As $\widetilde{V}_{\alpha\beta}$ is traceless symmetric, $v_{0}=v_{X}'=v_{Y}'=v_{Z}'=0$.

The number of active components in the above tensors is determined by the local site symmetry at the nuclear site.
The relation between the site symmetry and the active components are summarized in Table~\ref{tbl:activeq}, where the primary axis is taken as $Z$ axis.
Note that the symmetric off-diagonal components can be eliminated by appropriate rotation of the local coordinate frame, although the anti-symmetric components cannot be.
The number of the symmetric off-diagonal components is equivalent to that of the free angle parameters to diagonalize the symmetric part of the tensor.
Only the local symmetries 1, $\bar{1}$, 2, m, and 2/m allow the presence of symmetric off-diagonal components, and 1, $\bar{1}$, 2, m, 2/m, 4, $\bar{4}$, 4/m, 3, $\bar{3}$, 6, $\bar{6}$, and 6/m without vertical mirror allow the anti-symmetric off-diagonal components.

Once $\hat{K}_{\alpha\beta}^{(j)}$ and $\hat{\widetilde{V}}_{\alpha\beta}^{(j)}$ are determined at the particular site of a crystal (we assign $j=1$ for this representative site), the tensors of the other symmetry-equivalent sites can be obtained by the symmetry operation as $\hat{K}^{(j)}=\hat{R}_{j}\hat{K}^{(1)}\hat{R}_{j}^{-1}$, and $\hat{\widetilde{V}}_{\alpha\beta}^{(j)}$ is transformed in a similar way.
Here, $\hat{R}_{j}$ is the representation matrix of the polar vector without the (partial) translation, and the corresponding symmetry operation with the (partial) translation transforms the nuclear site from \#1 to \#$j$.
Note that when we consider the transformation property of the shift and EFG tensors, it is sufficient to take account of the symmetry operations without the (partial) translation of the associated point group.
The coefficient of the basis $\mathbb{X}_{i}$ for the shift tensor at $j$ site, $\hat{K}^{(j)}$, is explicitly obtained by
\begin{align}
&
k_{i}^{(j)}=\sum_{k}k_{k}^{(1)}{\rm Tr}(\mathbb{Q}_{i}\hat{R}_{j}\mathbb{Q}_{k}\hat{R}_{j}^{-1}),
\\&
k_{i}'^{(j)}=-\sum_{k}k_{k}'^{(1)}{\rm Tr}(\mathbb{G}_{i}\hat{R}_{j}\mathbb{G}_{k}\hat{R}_{j}^{-1}),
\end{align}
where common axes should be used for $\hat{R}$ and $\mathbb{Q}$, $\mathbb{G}$.

\section{Configuration determination}

\begin{figure}[t!]
\begin{center}
\includegraphics[width=0.8 \hsize ]{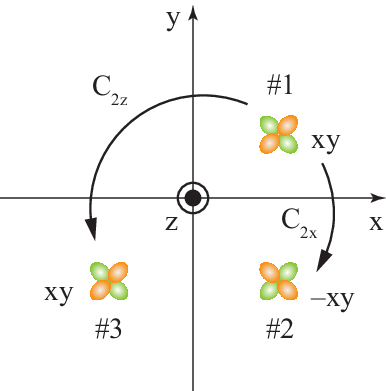} 
\caption{\label{fig:c2}
Pair of sites (\#1 and \#2) with the opposite-sign off-diagonal components ($xy$ and $zx$) related by $C_{2}$ rotation along $x$ axis ($C_{2x}$).
For $C_{2}$ rotation along $z$ axis ($C_{2z}$), $yz$ and $zx$ components of \#1 and \#3 sites constitute a pair.
}
\end{center}
\end{figure}

The local information at the nuclear site can be extracted by NMR measurement.
In particular, some of the off-diagonal components change their signs depending on the chirality or polarity of crystal, since certain component of the coordinate vector changes its sign by reversing the chirality or polarity of crystal.
Therefore, it may be useful to determine a configuration of chirality and/or polarity by analyzing the anisotropic NMR shift carefully, because its magnetic-field angle dependence reflects the sign and magnitude of the off-diagonal components.
In this section, we consider the condition to realize a configuration determination, and then we demonstrate a few example of the determination in the next section.

Let us first consider the property of the shift tensor at a general Wyckoff position, which has only identity symmetry operation~\cite{ITA16}.
At such a nuclear site, its shift tensor has all of the off-diagonal components as shown in Table~\ref{tbl:activeq}.

In the case of cubic system that always has three 2-fold axes being orthogonal with each other, there exists a pair of symmetry-equivalent sites having the off-diagonal components with the opposite signs, as shown in Fig.~\ref{fig:c2}.
Since the contribution of such a pair of sites to the NMR spectrum is identical to that of the corresponding pair in the crystal with opposite chirality or polarity, a configuration determination does not work out in cubic system.

Except for cubic system, the main crystallographic principal axis to characterize the chirality or polarity is $c$ axis (taken as $z$ axis) for hexagonal, trigonal, tetragonal, and orthorhombic systems, while $b$ axis (taken as $y$ axis) for monoclinic system in the standard setting.
When the chirality or polarity of crystal is reversed, the sign of $(yz,zx)$ components is reversed for the former systems, while that of $(yz,xy)$ components is reversed for the latter.
Meanwhile, there exists 2-fold $z$ axis for hexagonal, tetragonal, and orthorhombic systems.
Since there exists a pair of sites having the opposite signs of $(yz,zx)$ components as in the cubic system, it is impossible to perform a configuration determination of the chirality or polarity.

\begin{table*}
\caption{Sohncke space group including chiral one and the shift tensor parameters at general Wyckoff position (WP), where $z_{0}=1/3$.
Note that the relations between the crystal $(abc)$ and orthogonal $(xyz)$ axes are given by $a\parallel x$, $c\parallel z$ for No.151/153 and No.152/154.
The relations, $(xyz)=(YZX)$ for No.151/153 and $(xyz)=(ZXY)$ for No.152/154, are used, as $C_{2}$ rotations exist along $y$ and $x$ axes, respectively.
The signs in No.144/145, No.151/153, No.152/154 are for the right/left-handed structures, respectively.
The asymmetric magnetic-field angle dependence reflecting the chirality appears in the ``plane''.
Note that $k_{x}=k_{z}=k_{x}'=k_{z}'=0$ for the WPs (3a,3b) ($y_{b}=2x_{a}$, $z_{c}=-1/3,1/6$) in No.151/153, while, $k_{y}=k_{z}=k_{y}'=k_{z}'=0$ for (3a,3b) ($y_{b}=0$, $z_{c}=1/3,-1/6$) in No.152/154.
$(x_{a},y_{b},z_{c})$ is the fractional coordinate of primitive translation vectors.
}
\label{tbl:shiftchiral}
\centering{
\begin{tabular}{c|c|cc|ccccccccc|c} \hline\hline
Space group & WP & \# & Site & $\mathbb{Q}_{0}$ & $\mathbb{Q}_{u}$ & $\mathbb{Q}_{v}$ & $\mathbb{Q}_{yz}$ & $\mathbb{Q}_{zx}$ & $\mathbb{Q}_{xy}$ & $\mathbb{G}_{x}$ & $\mathbb{G}_{y}$ & $\mathbb{G}_{z}$ & Plane \\ \hline
149 (D$_{3}^{1}$, P312) & 6l(3j,3k) & & ($z_{0}=0$) & \multicolumn{3}{l}{(=No.$151$)} & & & & & & & $zx$ \\ \hline
150 (D$_{3}^{2}$, P321) & 6g(3e,3f) & & ($z_{0}=0$) & \multicolumn{3}{l}{(=No.$152$)} & & & & & & & $yz$ \\ \hline
151/153 & 6c(3a,3b) & 1(1) & $(x_{a},y_{b},\pm z_{c})$ & $k_{0}$ & $k_{u}$ & $k_{v}$ & $\pm k_{x}$ & $\pm k_{y}$ & $k_{z}$ & $\pm k_{x}'$ & $\pm k_{y}'$  & $k_{z}'$ & $zx$ \\
(D$_{3}^{3}$, P3$_{1}$12/D$_{3}^{5}$, P3$_{2}$12) &   & 2(1) & $(-x_{a}+y_{b},y_{b},\mp z_{c}\pm z_{0})$ & $k_{0}$ & $k_{u}$ & $k_v$ & $\mp k_x$ & $\pm k_y$ & $-k_z$ & $\mp k_x'$ & $\pm k_y'$ & $-k_z'$ & \\
  &    & 3(2) & $(x_{a},x_{a}-y_{b},\mp z_{c})$ & $k_{0}$ & $k_{u}$ & $k_{v}^{(+)}$ & $\mp k_{x}^{(+)}$ & $\pm k_{y}^{(-)}$ & $-k_{z}^{(-)}$ & $\mp k_{x}'^{(-)}$ & $\pm k_{y}^{(+)}$ & $-k_z'$ & \\
  &    & 4(3) & $(-y_{b},-x_{a},\mp z_{c}\mp z_{0})$ & $k_{0}$ & $k_{u}$ & $k_{v}^{(-)}$ & $\mp k_{x}^{(-)}$ & $\pm k_{y}^{(+)}$ & $-k_{z}^{(+)}$ & $\mp k_{x}'^{(+)}$ & $\pm k_{y}'^{(-)}$ & $-k_z'$ & \\
  &    & 5(3) & $(-y_{b},x_{a}-y_{b},\pm z_{c}\pm z_{0})$ & $k_{0}$ & $k_{u}$ & $k_{v}^{(+)}$ & $\pm k_{x}^{(+)}$ & $\pm k_{y}^{(-)}$ & $k_{z}^{(-)}$ & $\pm k_{x}'^{(-)}$ & $\pm k_{y}'^{(+)}$ & $k_z'$ & \\
  &    & 6(2) & $(-x_{a}+y_{b},-x_{a},\pm z_{c}\mp z_{0})$ & $k_{0}$ & $k_{u}$ & $k_{v}^{(-)}$ & $\pm k_{x}^{(-)}$ & $\pm k_{y}^{(+)}$ & $k_{z}^{(+)}$ & $\pm k_{x}'^{(+)}$ & $\pm k_{y}'^{(-)}$ & $k_z'$ & \\ \hline
152/154 & 6c(3a,3b) & 1(1) & $(x_{a},y_{b},\pm z_{c})$ & $k_{0}$ & $k_{u}$ & $k_{v}$ & $\pm k_{x}$ & $\pm k_{y}$ & $k_{z}$ & $\pm k_{x}'$ & $\pm k_{y}'$ & $k_{z}'$ & $yz$ \\
 (D$_{3}^{4}$, P3$_{1}$21/D$_{3}^{6}$, P3$_{2}$21) &   & 2(1) & $(x_{a}-y_{b},-y_{b},\mp z_{c}\mp z_{0})$ & $k_{0}$ & $k_{u}$ & $k_v$ & $\pm k_x$ & $\mp k_y$ & $-k_z$ & $\pm k_x'$ & $\mp k_y'$ & $-k_z'$ & \\
  &    & 3(2) & $(-x_{a},-x_{a}+y_{b},\mp z_{c}\pm z_{0})$ & $k_{0}$ & $k_{u}$ & $k_{v}^{(+)}$ & $\pm k_{x}^{(+)}$ & $\mp k_{y}^{(-)}$ & $-k_{z}^{(-)}$ & $\pm k_{x}'^{(-)}$ & $\mp k_{y}'^{(+)}$ & $-k_z'$ & \\
  &    & 4(3) & $(y_{b},x_{a},\mp z_{c})$ & $k_{0}$ & $k_{u}$ & $k_{v}^{(-)}$ & $\pm k_{x}^{(-)}$ & $\mp k_{y}^{(+)}$ & $-k_{z}^{(+)}$ & $\pm k_{x}'^{(+)}$ & $\mp k_{y}'^{(-)}$ & $-k_z'$ & \\
  &    & 5(3) & $(-y_{b},x_{a}-y_{b},\pm z_{c}\pm z_{0})$ & $k_{0}$ & $k_{u}$ & $k_{v}^{(+)}$ & $\pm k_{x}^{(+)}$ & $\pm k_{y}^{(-)}$ & $k_{z}^{(-)}$ & $\pm k_{x}'^{(-)}$ & $\pm k_{y}'^{(+)}$ & $k_z'$ & \\
  &    & 6(2) & $(-x_{a}+y_{b},-x_{a},\pm z_{c}\mp z_{0})$ & $k_{0}$ & $k_{u}$ & $k_{v}^{(-)}$ & $\pm k_{x}^{(-)}$ & $\pm k_{y}^{(+)}$ & $k_{z}^{(+)}$ & $\pm k_{x}'^{(+)}$ & $\pm k_{y}'^{(-)}$ & $k_z'$ & \\ \hline
155 (D$_{3}^{7}$, R32) & 18f(9d,9e) & & ($z_{0}=0$) & \multicolumn{3}{l}{(=No.$152$)} & & & & & & & $yz$ \\ \hline
\end{tabular}
}
\end{table*}

\begin{table*}
\caption{Polar space group and the shift tensor parameters at general Wyckoff position (WP), where $z_{0}=0$.
Note that $b\parallel y$ and $x$, $z$ are taken arbitrary for No.6, and $a\parallel x$, $c\parallel z$ for No.156 and No.157.
The relations, $y\parallel Z$ for No.6, $(xyz)=(ZXY)$ for No.156, and $(xyz)=(YZX)$ for No.157, are used.
The asymmetric magnetic-field angle dependence reflecting the polarity appears in the ``plane''.
Note that $k_{x}=k_{z}=k_{x}'=k_{z}'=0$ for the WPs (1a,1b) ($y_{b}=0,1/2$) in No.6 and (3c) ($y_{b}=0$) in No.157, while, $k_{y}=k_{z}=k_{y}'=k_{z}'=0$ for  (3d) ($y_{b}=2x_{a}$) in No.156.
}
\label{tbl:shiftpolar}
\centering{
\begin{tabular}{c|c|cc|ccccccccc|c} \hline\hline
Space group & WP & \# & Site & $\mathbb{Q}_{0}$ & $\mathbb{Q}_{u}$ & $\mathbb{Q}_{v}$ & $\mathbb{Q}_{yz}$ & $\mathbb{Q}_{zx}$ & $\mathbb{Q}_{xy}$ & $\mathbb{G}_{x}$ & $\mathbb{G}_{y}$ & $\mathbb{G}_{z}$ & Plane \\ \hline
6 (C$_{\rm s}^{1}$, Pm) & 2c(1a,1b) & 1(1) & $(x_{a},y_{b},z_{c})$ & $k_{0}$ & $k_{u}$ & $k_{v}$ & $k_{x}$ & $k_{y}$ & $k_{z}$ & $k_{x}'$ & $k_{y}'$ & $k_{z}'$ & $zx$ \\
& & 2(1) & $(x_{a},-y_{b},z_{c}+z_{0})$ & $k_{0}$ & $k_{u}$ & $k_v$ & $-k_x$ & $k_y$ & $-k_z$ & $-k_x'$ & $k_y'$ & $-k_z'$ \\ \hline
7 (C$_{\rm s}^{2}$, Pc) & 2a & & ($z_{0}=1/2$) & \multicolumn{3}{l}{(=No.$6$)} & & & & & & & $zx$ \\ \hline
8 (C$_{\rm s}^{3}$, Cm) & 4b & & ($z_{0}=0$) & \multicolumn{3}{l}{(=No.$6$)} & & & & & & & $zx$ \\ \hline
9 (C$_{\rm s}^{4}$, Cc) & 4a & & ($z_{0}=1/2$) & \multicolumn{3}{l}{(=No.$6$)} & & & & & & & $zx$ \\ \hline
156 (C$_{\rm 3v}^{1}$, P3m1) & 6e(3d) & 1(1) & $(x_{a},y_{b},z_{c})$ & $k_{0}$ & $k_{u}$ & $k_{v}$ & $k_{x}$ & $k_{y}$ & $k_{z}$ & $k_{x}'$ & $k_{y}'$ & $k_{z}'$ & $yz$ \\
& & 2(2) & $(-y_{b},x_{a}-y_{b},z_{c})$ & $k_{0}$ & $k_{u}$ & $k_{v}^{(+)}$ & $k_{x}^{(+)}$ & $k_{y}^{(-)}$ & $k_{z}^{(-)}$ & $k_{x}'^{(-)}$ & $k_{y}'^{(+)}$ & $k_{z}'$ & \\
& & 3(3) & $(-x_{a}+y_{b},-x_{a},z_{c})$ & $k_{0}$ & $k_{u}$ & $k_{v}^{(-)}$ & $k_{x}^{(-)}$ & $k_{y}^{(+)}$ & $k_{z}^{(+)}$ & $k_{x}'^{(+)}$ & $k_{y}'^{(-)}$ & $k_{z}'$ & \\
& & 4(1) & $(-x_{a}+y_{b},y_{b},z_{c}+z_{0})$ & $k_{0}$ & $k_{u}$ & $k_v$ & $k_x$ & $-k_y$ & $-k_z$ & $k_x'$ & $-k_y'$ & $-k_z'$ & \\
& & 5(3) & $(x_{a},x_{a}-y_{b},z_{c}+z_{0})$ & $k_{0}$ & $k_{u}$ & $k_{v}^{(+)}$ & $k_{x}^{(+)}$ & $-k_{y}^{(-)}$ & $-k_{z}^{(-)}$ & $k_{x}'^{(-)}$ & $-k_{y}'^{(+)}$ & $-k_{z}'$ & \\
& & 6(2) & $(-y_{b},-x_{a},z_{c}+z_{0})$ & $k_{0}$ & $k_{u}$ & $k_{v}^{(-)}$ & $k_{x}^{(-)}$ & $-k_{y}^{(+)}$ & $-k_{z}^{(+)}$ & $k_{x}'^{(+)}$ & $-k_{y}'^{(-)}$ & $-k_{z}'$ & \\ \hline
157 (C$_{\rm 3v}^{2}$, P31m) & 6d(3c) & 1(1) & $(x_{a},y_{b},z_{c})$ & $k_{0}$ & $k_{u}$ & $k_{v}$ & $k_{x}$ & $k_{y}$ & $k_{z}$ & $k_{x}'$ & $k_{y}'$ & $k_{z}'$ & $zx$ \\
& & 2(2) & $(-y_{b},x_{a}-y_{b},z_{c})$ & $k_{0}$ & $k_{u}$ & $k_{v}^{(+)}$ & $k_{x}^{(+)}$ & $k_{y}^{(-)}$ & $k_{z}^{(-)}$ & $k_{x}'^{(-)}$ & $k_{y}'^{(+)}$ & $k_{z}'$ & \\
& & 3(3) & $(-x_{a}+y_{b},-x_{a},z_{c})$ & $k_{0}$ & $k_{u}$ & $k_{v}^{(-)}$ & $k_{x}^{(-)}$ & $k_{y}^{(+)}$ & $k_{z}^{(+)}$ & $k_{x}'^{(+)}$ & $k_{y}'^{(-)}$ & $k_{z}'$ & \\
& & 4(1) & $(x_{a}-y_{b},-y_{b},z_{c}+z_{0})$ & $k_{0}$ & $k_{u}$ & $k_v$ & $-k_x$ & $k_y$ & $-k_z$ & $-k_x'$ & $k_y'$ & $-k_z'$ & \\
& & 5(3) & $(-x_{a},-x_{a}+y_{b},z_{c}+z_{0})$ & $k_{0}$ & $k_{u}$ & $k_{v}^{(+)}$ & $-k_{x}^{(+)}$ & $k_{y}^{(-)}$ & $-k_{z}^{(-)}$ & $-k_{x}'^{(-)}$ & $k_{y}'^{(+)}$ & $-k_{z}'$ & \\
& & 6(2) & $(y_{b},x_{a},z_{c}+z_{0})$ & $k_{0}$ & $k_{u}$ & $k_{v}^{(-)}$ & $-k_{x}^{(-)}$ & $k_{y}^{(+)}$ & $-k_{z}^{(+)}$ & $-k_{x}'^{(+)}$ & $k_{y}'^{(-)}$ & $-k_{z}'$ & \\ \hline
158 (C$_{\rm 3v}^{3}$, P3c1) & 6d & & ($z_{0}=1/2$) & \multicolumn{3}{l}{(=No.$156$)} & & & & & & & $yz$ \\ \hline
159 (C$_{\rm 3v}^{4}$, P31c) & 6c & & ($z_{0}=1/2$) & \multicolumn{3}{l}{(=No.$157$)} & & & & & & & $zx$ \\ \hline
160 (C$_{\rm 3v}^{5}$, R3m) & 18c(9b) & & ($z_{0}=0$) & \multicolumn{3}{l}{(=No.$156$)} & & & & & & & $yz$ \\ \hline
161 (C$_{\rm 3v}^{6}$, R3c) & 18b & & ($z_{0}=1/2$) & \multicolumn{3}{l}{(=No.$156$)} & & & & & & & $yz$ \\ \hline\hline
\end{tabular}
}
\end{table*}

\begin{table*}
\caption{Sohncke and polar space group and the shift tensor parameters at general Wyckoff position (WP), where $y_{0}=0$ and $z_{0}=1/3$.
Note that $b\parallel y$ and $x$, $z$ are taken arbitrary for No.3, and $a\parallel x$, $c\parallel z$ for No.144, and $x$, $y$ $z$ are taken arbitrary for No.1.
The relation, $y\parallel Z$ for No.3, is used.
The sign in No.144/145 are for the right/left-handed crystals, respectively.
The asymmetric magnetic-field angle dependence reflecting the chirality and polarity appears in the ``plane''.
Note that $k_{x}=k_{z}=k_{x}'=k_{z}'=0$ for the WPs (1a,1b,1c,1d) ($x_{a}=0,0,1/2,1/2$, $z_{c}=0,1/2,0,1/2$) in No.3.
}
\label{tbl:shiftpc}
\centering{
\begin{tabular}{c|c|cc|ccccccccc|c} \hline\hline
Space group & WP & \# & Site & $\mathbb{Q}_{0}$ & $\mathbb{Q}_{u}$ & $\mathbb{Q}_{v}$ & $\mathbb{Q}_{yz}$ & $\mathbb{Q}_{zx}$ & $\mathbb{Q}_{xy}$ & $\mathbb{G}_{x}$ & $\mathbb{G}_{y}$ & $\mathbb{G}_{z}$ & Plane \\ \hline
1 (C$_{1}^{1}$, P1) & 1a & 1 & $(x_{a},y_{b},z_{c})$ & $k_{0}$ & $k_{u}$ & $k_{v}$ & $k_{x}$ & $k_{y}$ & $k_{z}$ & $k_{x}'$ & $k_{y}'$ & $k_{z}'$ & all \\ \hline
3 (C$_{2}^{1}$, P2) & 2e(1a,1b,1c,1d) & 1(1) & $(x_{a},y_{b},z_{c})$ & $k_{0}$ & $k_{u}$ & $k_{v}$ & $k_{x}$ & $k_{y}$ & $k_{z}$ & $k_{x}'$ & $k_{y}'$ & $k_{z}'$ & $zx$ \\
  & & 2(1) & $(-x_{a},y_{b}+y_{0},-z_{c})$ & $k_{0}$ & $k_{u}$ & $k_v$ & $-k_x$ & $k_y$ & $-k_z$ & $-k_x'$ & $k_y'$ & $-k_z'$ & \\ \hline
4 (C$_{2}^{2}$, P2$_{1}$) & 2a & & ($y_{0}=1/2$) & \multicolumn{3}{l}{(=No.$3$)} & & & & & & & $zx$ \\ \hline
5 (C$_{2}^{3}$, C2) & 4c(2a,2b) & & ($y_{0}=0$) & \multicolumn{3}{l}{(=No.$3$)} & & & & & & & $zx$ \\ \hline
143 (C$_{3}^{1}$, P3) & 3d & & ($z_{0}=0$) &  \multicolumn{3}{l}{(=No.$144$)} & & & & & & & all \\ \hline
144/145 & 3a & 1 & $(x_{a},y_{b},\pm z_{c})$ & $k_{0}$ & $k_{u}$ & $k_{v}$ & $\pm k_{x}$ & $\pm k_{y}$ & $k_{z}$ & $\pm k_{x}'$ & $\pm k_{y}'$ & $k_{z}'$ & all \\
(C$_{3}^{2}$, P3$_{1}$/C$_{3}^{3}$, P3$_{2}$) & & 2 & $(-y_{b},x_{a}-y_{b},\pm z_{c}\pm z_{0})$ & $k_{0}$ & $k_{u}$ & $k_{v}^{(+)}$ & $\pm k_{x}^{(+)}$ & $\pm k_{y}^{(-)}$ & $k_{z}^{(-)}$ & $\pm k_{x}'^{(-)}$ & $\pm k_{y}'^{(+)}$ & $k_z'$ & \\
&    & 3 & $(-x_{a}+y_{b},-x_{a},\pm z_{c}\mp z_{0})$ & $k_{0}$ & $k_{u}$ & $k_{v}^{(-)}$ & $\pm k_{x}^{(-)}$ & $\pm k_{y}^{(+)}$ & $k_{z}^{(+)}$ & $\pm k_{x}'^{(+)}$ & $\pm k_{y}'^{(-)}$ & $k_z'$ & \\ \hline
146 (C$_{3}^{4}$, R3) & 9b & & ($z_{0}=0$) & \multicolumn{3}{l}{(=No.$144$)} & & & & & & & all \\ \hline
\end{tabular}
}
\end{table*}

With these observations, a configuration determination of chirality or polarity is possible only for trigonal (C$_{3}$, D$_{3}$, C$_{\rm 3v}$), monoclinic (C$_{2}$, C$_{\rm s}$), and triclinic (C$_{1}$) systems.
Among these groups, D$_{3}$ is the chiral point group, C$_{\rm s}$, C$_{\rm 3v}$ are the polar point groups, and C$_{1}$, C$_{2}$, C$_{3}$ are the chiral and polar point groups.
The corresponding space groups are Nos.149--155 (D$_{3}$), Nos.6--9 (C$_{\rm s}$), Nos.156--161 (C$_{\rm 3v}$), No.1 (C$_{1}$), Nos.3--5 (C$_{2}$), and Nos.143--146 (C$_{3}$), respectively.

The above conclusion is applicable to the Wyckoff positions with higher symmetry within the above space groups as long as their site symmetry belongs to 2 or m as shown in Table~\ref{tbl:activeq}.
We summarize all the relevant space groups and the site dependences of their expansion coefficients in Table~\ref{tbl:shiftchiral} for Sohncke space group including chiral one, in Table~\ref{tbl:shiftpolar} for polar space group, and in Table~\ref{tbl:shiftpc} for Sohncke and polar space group.
Note that the Tables are expressed in terms of the crystallographic orthogonal coordinate $(xyz)$ by taking the appropriate correspondences between $(xyz)$ and $(XYZ)$.
$k_{0}$ and $k_{u}$ are common in all sites.
We have introduced the following abbreviations arising from 3-fold rotation for notational simplicity,
\begin{align}
&
k_{v}^{(\pm)}=\frac{-k_{v}\pm\sqrt{3}k_{z}}{2},
\quad
k_{z}^{(\pm)}=\frac{-k_{z}\pm\sqrt{3}k_{v}}{2},
\cr&
k_{x}^{(\pm)}=\frac{-k_{x}\pm\sqrt{3}k_{y}}{2},
\quad
k_{y}^{(\pm)}=\frac{-k_{y}\pm\sqrt{3}k_{x}}{2},
\cr&
k_{x}'^{(\pm)}=\frac{-k_{x}'\pm\sqrt{3}k_{y}'}{2},
\quad
k_{y}'^{(\pm)}=\frac{-k_{y}'\pm\sqrt{3}k_{x}'}{2}.
\end{align}

\section{Examples}

\begin{figure}[t!]
\begin{center}
\includegraphics[width=1.0\hsize ]{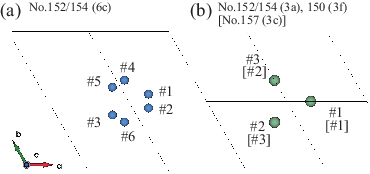} 
\caption{\label{fig:WP}
Equivalent Wyckoff positions of the trigonal space groups: (a) 6c for No.152/154, (b) 3a, 3f, 3c for No.152/154, No.150, No.157, respectively.
}
\end{center}
\end{figure}

In this section, we present three prime examples for the configuration determination of chirality or polarity: (A) the simple chiral system, Te, (B) the chiral system IrSn$_{4}$ and RhSn$_{4}$ with general Wyckoff position, and (C) URhSn that is considered as a chiral or polar system brought about by the spontaneous symmetry breaking.
In all these example, the nuclear spin of Te or Sn is $I=1/2$ and the quadrupolar interaction is absent.
As the hyperfine interaction is sufficiently small or assumed to be small for URhSn as compared to $\hbar\omega_{0}$, we treat the NMR shift in the first-order perturbation (\ref{eq:shiftpb}), and $k_{x}'$, $k_{y}'$, and $k_{z}'$ can be omitted.
The applied magnetic-field direction is expressed by two angles ($\theta$, $\phi$) as
\begin{align}
\bm{h}=(\sin\theta\cos\phi,\sin\theta\sin\phi,\cos\theta),
\end{align}
in the $(xyz)$ coordinate.

\subsection{Te}

Let us first consider the case of elemental Te~\cite{Koma73,Koma73a,Furukawa17,Furukawa21,Oiwa22}.
The space group of Te is No.152/154 (right/left-handed structure), and $^{125}$Te is located at Wyckoff 3a position as shown in Fig.~\ref{fig:WP}(b), and its natural abundance (NA) is $7.07\%$ and $\gamma=13.4327$ MHz/T.
As the system has $C_{2}$ rotation along $x$ axis, $C_{2x}$, at \#1 site, $k_{y}=k_{z}=0$ as shown in Table~\ref{tbl:shiftchiral}.
Namely, the main crystallographic principal axis $Z$ is regarded as $x\parallel a$ in crystallographic orthogonal coordinate.

The shift tensor for the right-handed structure were determined experimentally~\cite{Koma73} as $\sigma_{zz}=8.860\times 10^{-4}$, $\sigma_{xx}=\sigma_{zz}-8.572\times 10^{-4}$, $\sigma_{yy}=\sigma_{zz}+1.146\times 10^{-4}$, $\sigma_{yz}=\sigma_{zy}=-8.830\times 10^{-4}$. By considering the different convention of shift tensors, i.e., $\hat{K}^{(1)}=-\hat{\sigma}$, the expansion coefficients are
\begin{align}
&
k_{0}^{(1)}=-1.1059\times 10^{-3},
\quad
k_{u}^{(1)}=-3.0317\times 10^{-4},
\cr&
k_{v}^{(1)}=6.8717\times 10^{-4},
\quad
k_{x}^{(1)}=1.2488\times 10^{-3}.
\label{eq:teval}
\end{align}
The parameters for the left-handed structure (No.154) can be obtained by the mirror operation with respect to $xy$ plane, which gives the sign change as $k_{x}^{(1)}\to -k_{x}^{(1)}$.
Note that the left-handed structure is also obtained by the mirror operation with respect to $zx$ or its equivalent planes.

\begin{figure}[t!]
\begin{center}
\includegraphics[width=0.9\hsize ]{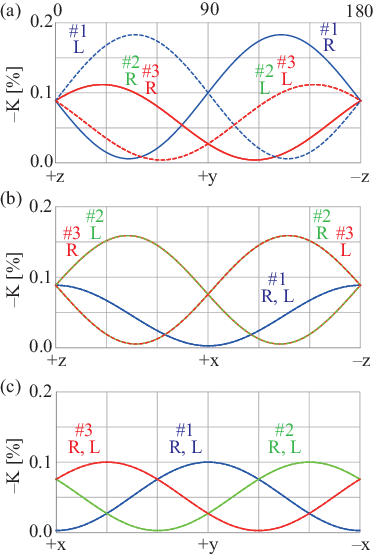} 
\caption{\label{fig:Te}
Field-angle dependences of the NMR shift of $^{125}$Te at 3a Wykoff position for (a) $yz$ plane, (b) $zx$ plane, and (c) $xy$ plane, where the solid (dashed) lines represent for the right (left) handed structure.
The shift is asymmetric in (a), and the lines of the right and left handed structure are related with each other by the reflection with respect to $y$ axis.
It is impossible to determine a configuration of chirality from (b) and (c).
}
\end{center}
\end{figure}

The field-angle dependences of the NMR shift calculated with (\ref{eq:teval}) are shown in Fig.~\ref{fig:Te} for (a) $yz$ plane, (b) $zx$ plane, and (c) $xy$ plane.
These dependences were already discussed in the pioneering work by Koma~\cite{Koma73}.
As expected from the column ``Plane'' in Table~\ref{tbl:shiftchiral}, the difference between the opposite chirality appears in $yz$ plane.
The solid lines for the right-handed structure is asymmetric, and these lines are the reflection of those for the left-handed structure (the dashed lines) with respect to $y$ axis.
This comes from the fact that there is no pairs of the sites with opposite signs of the coefficients $k_{x}$, $k_{x}^{(+)}$, and $k_{x}^{(-)}$ for one handedness, and chirality reverses all of their signs.
See the column $\mathbb{Q}_{yz}$ of Table~\ref{tbl:shiftchiral}.
Indeed, the following relation holds between the NMR shift of the right-handed (R) and left-handed (L) structures,
\begin{align}
K^{(j)}_{\rm R}(\theta,\phi=\pi/2)=K^{(j)}_{\rm L}(\pi-\theta,\phi=\pi/2).
\label{eq:lrshift}
\end{align}
Note that \#2 and \#3 sites are equivalent in this plane owing to the presence of $C_{2x}$ rotation.

On the other hand, in $zx$ plane in Fig.~\ref{fig:Te}(b), the contributions of the pair of \#2 and \#3 sites for each chirality are symmetric with respect to $x$ axis, even though each contribution shows asymmetric angular dependence.
This is because \#2 and \#3 sites are related with each other by the $C_{2x}$ rotation as discussed in the previous section, which leads to the sign reversal of $k_{y}^{(\pm)}$ between \#2 and \#3 sites as shown in the column $\mathbb{Q}_{zx}$ of Table~\ref{tbl:shiftchiral}.
The field-angle dependences for both handedness coincide in this plane, and it is impossible to determine the configuration of the chirality in this plane.
Moreover, in $xy$ plane, there is no distinction between both handedness.

The anti-symmetric part of the angle dependence in $yz$ plane can be extracted by
\begin{align}
A(yz)&\equiv\frac{1}{2}[K^{(1)}(\theta,\phi=\pi/2)-K^{(1)}(\pi-\theta,\phi=\pi/2)]
\cr&
=2K_{yz}\sin(2\theta).
\label{eq:antisym}
\end{align}
The sign and magnitude of the off-diagonal components $K_{yz}$ can be determined by the linear dependence in $A(yz)$ around $\theta=\pi/2$.

\begin{figure}[t!]
\begin{center}
\includegraphics[width=0.9\hsize ]{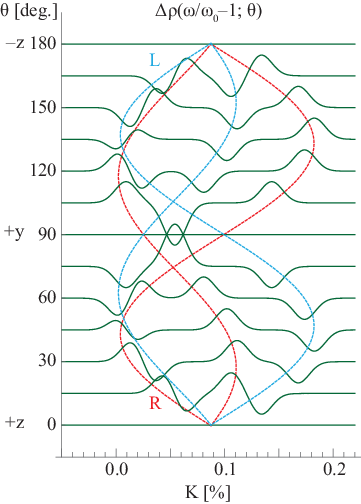} 
\caption{\label{fig:spectrum}
Subtracted spectrum (\ref{eq:subtspec}) for Te ($w_{\rm R}>w_{\rm L}$) in arbitrary unit.
Gaussian profiles with the second moment $\Delta K=0.015$ \% are used for each resonance line.
The positive (negative) intensity represents the contribution of the right (left) handed domain.
The red and blue dotted lines represent the angle dependence of the NMR shift in the right and left handed domains, respectively.
}
\end{center}
\end{figure}

When both handedness is mixed in the presence of twin domains, the contributions of both handedness appear in the spectrum.
By means of the relation (\ref{eq:lrshift}), the contributions of the right and left handed structures can be decomposed by the subtraction of the observed spectrum,
\begin{align}
&
\rho(\omega;\theta)=w_{\rm R}\bar{\rho}_{\rm R}(\omega;\theta)+w_{\rm L}\bar{\rho}_{\rm L}(\omega;\theta),
\cr&\quad\quad\quad
\bar{\rho}_{\rm R}(\omega;\theta)=\bar{\rho}_{\rm L}(\omega;\pi-\theta),
\label{eq:rho}
\end{align}
as
\begin{align}
\Delta\rho(\omega;\theta)&\equiv \rho(\omega;\theta)-\rho(\omega;\pi-\theta)
\cr&
=(w_{\rm R}-w_{\rm L})[\bar{\rho}_{\rm R}(\omega;\theta)-\bar{\rho}_{\rm L}(\omega;\theta)],
\label{eq:subtspec}
\end{align}
where $w_{\rm R,L}$ is the domain weight of the handedness, and $\bar{\rho}_{\rm R,L}(\omega;\theta)$ is an ideal spectrum from pure single domain.
Note that the subtracted spectrum vanishes when $w_{\rm R}=w_{\rm L}$, otherwise the positive (negative) intensity represents the right (left) handed spectrum for $w_{\rm R}>w_{\rm L}$.
An example plot of the subtracted spectrum for $w_{\rm R}>w_{\rm L}$ is given in Fig.~\ref{fig:spectrum}.
It may be useful to assign the branches arising from each handedness, and then the domain weight is determined by (\ref{eq:rho}).

As discussed above, it is possible to determine the relative ratio of the chirality of crystal by analyzing the field-angle dependence of the NMR shift.
However, in practice, the \textit{absolute} determination of the chirality is impossible by the NMR shift alone, because we cannot know a priori the sign of the off-diagonal components for given handedness.
For the absolute-configuration determination, one needs further input from the shape of the asymmetric etch pit on the cleaved face of crystal~\cite{Koma73,Furukawa21} or theoretical estimation of the sign of the off-diagonal components based on, e.g., the orbital current model for nonmagnetic crystals~\cite{Koma73a,Shimizu69,Xue00,Abragam61} or the dipole-dipole interaction for magnetic ones~\cite{Abragam61}.

\subsection{IrSn$_{4}$ and RhSn$_{4}$}

\begin{figure}[t!]
\begin{center}
\includegraphics[width=0.9\hsize ]{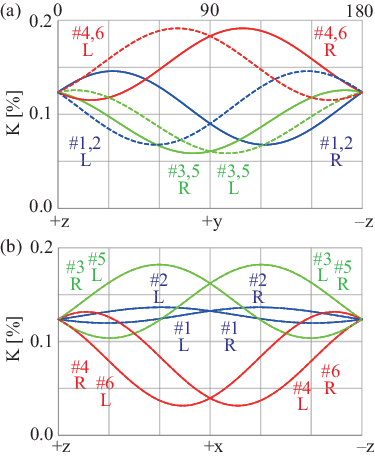} 
\caption{\label{fig:6c}
Field-angle dependences of the NMR shift of $^{117}$Sn at 6c Wykoff position in (a) $yz$ plane and (b) $zx$ plane.
}
\end{center}
\end{figure}

Next, we consider the case of a general Wyckoff position.
For such purpose, we take IrSn$_{4}$ and RhSn$_{4}$~\cite{Nakamura23} whose space group are the same as Te.
The ${}^{117}$Sn nucleus is located at 6c site as shown in Fig.~\ref{fig:WP}(a) [$(x,y,z)\simeq (0.23,0.5,0.43)$], and its NA is $7.5\%$.
Note that Ir and Rh are at 3a site, and the similar analysis of Te can be made.
Since the shift parameters are undetermined experimentally at the moment, we use the tentative values, $(k_{0}^{(1)},k_{u}^{(1)},k_{v}^{(1)},k_{x}^{(1)},k_{y}^{(1)},k_{z}^{(1)})=(20,1.0,3.0,5.0,1.0,10)\times 10^{-4}$.
It should be emphasized that the qualitative conclusions are not altered by different choices of parameters.

The field-angle dependences of the NMR shift are shown in Fig.~\ref{fig:6c} for (a) $yz$ plane and (b) $zx$ plane.
As similar to the results of Te, the difference between the right and left handed structures appears only in $yz$ plane.
Note that the pairs of sites, (\#1,\#2), (\#3,\#5), and (\#4,\#6), are equivalent in $yz$ plane as they are related by $C_{2x}$ rotation.
The same relations (\ref{eq:lrshift}) and (\ref{eq:antisym}) also hold for 6c site, and the analysis using (\ref{eq:subtspec}) can also be applied to decompose the contributions of twin domains.

\subsection{URhSn}

\begin{table}
\caption{
Wyckoff positions and site symmetry for URhSn.
}
\label{tbl:urhsn}
\centering{
\begin{tabular}{lllll} \hline\hline
Atom & Site (No.189) & No.189 & No.150 & No.157 \\ \hline
U & $(0.5925,0,0)$ & 3f (m2m) & 3e (.2.) & 3c (..m) \\
Rh1 & $(1/3,2/3,1/2)$ & 2d ($\bar{6}..$) & 2d (3..) & 2b (3..) \\
Rh2 & $(0,0,0)$ & 1a ($\bar{6}$2m) & 1a (32.) & 1a (31m) \\
Sn & $(1/4,0,1/2)$ & 3g (m2m) & 3f (.2.) & 3c (..m) \\ \hline\hline
\end{tabular}
}
\end{table}

\begin{figure*}[t!]
\begin{center}
\includegraphics[width=0.7\hsize ]{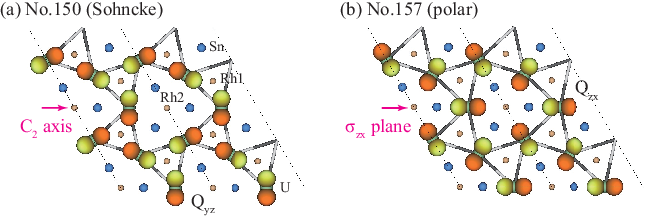} 
\caption{\label{fig:two_phase}
Two candidates of quadrupole order in URhSn, (a) $Q_{yz}$ order (No.150, Sohncke) and (b) $Q_{zx}$ order (No.157 polar).
}
\end{center}
\end{figure*}

\begin{figure*}[t!]
\begin{center}
\includegraphics[width=0.75\hsize ]{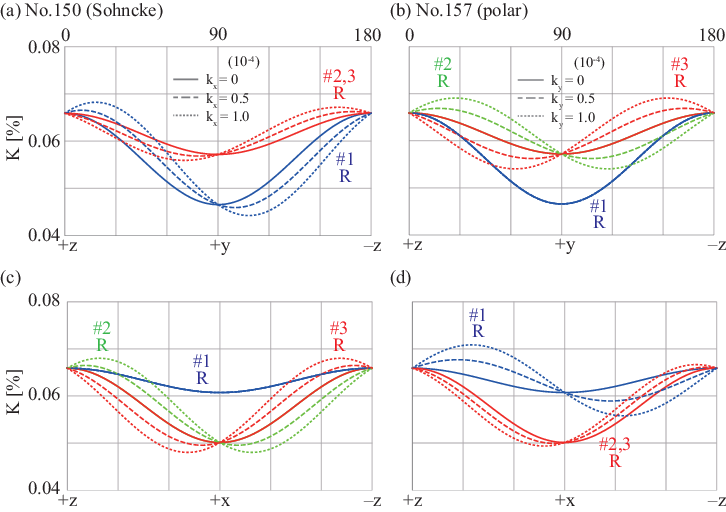} 
\caption{\label{fig:no150_157}
Field-angle dependences of the NMR shift of $^{117}$Sn for (a), (c) No.150 at 3f site and (b), (d) No.157 at 3c site.
Only the angle dependences for the right-handed structure are shown for clarity.
Note that $k_{x}$ ($k_{y}$) increases with increase of the order parameter in No.150 Sohncke (No.157 polar) phase.
}
\end{center}
\end{figure*}

The hexagonal compound URhSn belongs to the achiral nonpolar space group, No.189 (D$_{3h}^{3}$, P$\bar{6}$2m), and exhibits successive phase transitions at $T_{0}=54$ K and $T_{c}=16$ K~\cite{Shimizu20}.
The low-temperature phase below $T_{c}$ is ferromagnetic with the ordered moments along $c$ axis, while the intermediate phase between $T_{0}$ and $T_{c}$ is not clearly identified yet.
At the onset of the intermediate phase, the magnetic susceptibility shows only a weak anomaly and the structural deformation was not reported within the experimental accuracy~\cite{Miranbet95,Kruk97}.
It has been argued that the electric multipole order with the uniform ordering vector $\bm{q}=0$ is most likely~\cite{Ishitobi23,Harima23,Tabata23,Tokunaga23,Yanagisawa23}.

The recent resonant X-ray scattering experiment observed slight increase of the intensity only in the $\pi$-$\sigma'$ scattering, which suggests that the electric charge distribution has $z$ component, i.e., the quadrupole moment $Q_{yz}$ or $Q_{zx}$~\cite{Tabata23}.
Moreover, $C_{44}$ mode in the ultrasonic measurements shows relatively large softening toward $T_{0}$, which indicates that $Q_{yz}$ and $Q_{zx}$ are relevant in the intermediate phase~\cite{Yanagisawa23}.
The longitudinal $C_{33}$ mode also exhibits almost discontinuous softening just below $T_{0}$, which is consistent with the fact that Rh1 site position changes from $(1/3,2/3,1/2)$ to $(1/3,2/3,z)$ with the free parameter $z\approx 1/2$.
The preliminary result of Sn-NMR measurement suggests the lack of the $zx$ mirror plane~\cite{Tokunaga23}.

The ordering of the cluster of $Q_{yz}$ quadrupoles with A$_{1}''$ irreducible representation (irrep.) as shown in Fig.~\ref{fig:two_phase}(a) leads to symmetry lowering to No.150 (Sohncke group), while that of the cluster of $Q_{zx}$ quadrupoles with A$_{2}''$ irrep. (Fig.~\ref{fig:two_phase}(b)) leads to No.157 (polar group), both of which belong to the maximal subgroups of No.189~\cite{Ishitobi23,Harima23}.
The change of site symmetry due to these phase transitions is summarized in Table~\ref{tbl:urhsn}.

Although the NMR shift of $^{117}$Sn is symmetric in $yz$ and $zx$ planes with the extrema at $x$, $y$, and $z$ axes in the paramagnetic phase, it becomes asymmetric when either $Q_{yz}$ (No.150, Sohncke) or $Q_{zx}$ (No.157, polar) ordering is realized owing to the appearance of nonzero off-diagonal component as shown in Tables~\ref{tbl:shiftchiral} and \ref{tbl:shiftpolar}.
It should be emphasized that the asymmetric field-angle dependence appears in $yz$ plane for No.150 phase, while in $zx$ plane for No.157 phase as shown below.

At \#1 site, $k_{y}^{(1)}=k_{z}^{(1)}=0$ ($k_{x}^{(1)}=k_{z}^{(1)}=0$) owing to the presence of $C_{2x}$ rotation ($\sigma_{zx}$ mirror).
By using the tentative values, $(k_{0}^{(1)},k_{u}^{(1)},k_{v}^{(1)})=(10, 1.0, 1.0)\times 10^{-4}$, the field-angle dependences of the NMR shift in $yz$ and $zx$ planes are shown in Fig.~\ref{fig:no150_157} (a) for No.150 with $k_{x}^{(1)}=(0,0.5,1.0)\times 10^{-4}$ and (b) for No.157 with $k_{y}^{(1)}=(0,0.5,1.0)\times 10^{-4}$.
The contributions of \#2 and \#3 sites in $yz$ ($zx$) plane for No.150 (No.157) are equivalent as they are related by $C_{2x}$ rotation ($\sigma_{zx}$ mirror).
Note that the field-angle dependences of No.150 (No.157) for the left handed (negative polarity) structure are given by $k_{x}^{(1)}\to -k_{x}^{(1)}$ ($k_{y}^{(1)}\to -k_{y}^{(1)}$), and the similar relation (\ref{eq:lrshift}) between opposite chirality (polarity) also holds.
Note that the binary nature of chirality or polarity corresponds to the existence of opposite domain of the quadrupole order.
On the other hand, the contributions of the \#2 and \#3 sites are symmetric in $zx$ plane ($yz$ plane) for No.150 (No.157) since \#2 and \#3 sites are related with each other by the $C_{2x}$ rotation ($\sigma_{zx}$ mirror).

As the symmetry property at the Sn site is the same as elemental Te for No.150, the anti-symmetric part is also characterized by (\ref{eq:antisym}), while that for No.157 in $zx$ plane is given by
\begin{align}
A(zx)&\equiv\frac{1}{2}[K^{(1)}(\theta,\phi=0)-K^{(1)}(\pi-\theta,\phi=0)]
\cr&
=2K_{zx}\sin(2\theta).
\label{eq:antisymzx}
\end{align}
The anti-symmetric parts in (\ref{eq:antisym}) and (\ref{eq:antisymzx}) are proportional to $K_{yz}$ and $K_{zx}$, respectively, which are expected to increase with the growth of the order parameters, $\braket{Q_{yz}}$ and $\braket{Q_{zx}}$.
Therefore, they are expected to rapidly develop just below $T_{0}$.

By the same analysis using the subtracted spectrum as (\ref{eq:subtspec}), each contribution of the left or right handedness (positive or negative polarity) can be extracted.
Therefore, careful analysis of the NMR shift of $^{117}$Sn is useful to reveal the enigmatic phase transition at $T_{0}$ in URhSn.

\section{Summary}

We have investigated the condition to perform a configuration determination for chiral and/or polar crystals by using the asymmetric magnetic-field angle dependence of NMR shift in the appropriate plane, where the asymmetry depends on its chirality or polarity.
This method is employed for the triclinic, monoclinic, and trigonal crystal structures among 230 crystallographic space groups.
The appropriate plane to extract chirality or polarity from the asymmetric field-angle dependence is the plane perpendicular to the 2-fold rotation axis or the mirror plane of the crystal if it exists, otherwise all planes can be used.

We have discussed the asymmetric behaviors of the NMR shift in the appropriate plane, and how to identify each contribution of twin domains, and so on, by using the specific examples of Te, IrSn$_{4}$ and RhSn$_{4}$, and the spontaneous symmetry-breaking phases of URhSn.
The analysis of the anisotropic NMR shift provides a complementary method of a configuration determination for chirality or polarity, and quantifies the ratio of twin domains, which could stimulate further investigation on nonreciprocal phenomena and cross-correlated responses in the inversion-symmetry breaking systems.

\begin{acknowledgments}
We acknowledge many useful discussions with Yoshihiko Togawa, Hiroshi M. Yamamoto, Jun-ichiro Kishine, Yusuke Kato, Hidekazu Mukuda, Tetsuaki Itou, Masashi Takigawa, Tatsuma D. Matsuda, Kazumasa Hattori, Takayuki Ishitobi, Tatsuya Yanagisawa, Yo Tokunaga, and Chihiro Tabata.
This research was supported by JSPS KAKENHI Grants Numbers JP23K03288, JP23H00091, and the grants of Special Project (IMS program 23IMS1101), and OML Project (NINS program No, OML012301) by the National Institutes of Natural Sciences.
\end{acknowledgments}

\appendix

\section{On shift and EFG tensors}

\subsection{Hyperfine coupling and shift tensor}

Let us start with the hyperfine interaction at the $j$ nuclear site with electronic spin $\bm{S}_{k}$ (dimensionless) at $k$ site~\cite{Abragam61},
\begin{align}
H_{\rm hf}^{(j)}=\sum_{k}\sum_{\alpha\beta}^{x,y,z}A_{jk}^{\alpha\beta}I_{j}^{\alpha}S_{k}^{\beta},
\end{align}
where $A_{jk}^{\alpha\beta}$ is the hyperfine coupling tensor between $j$ and $k$ sites, which is second-rank polar tensor, and in general asymmetric.
The components of $A_{jk}^{\alpha\beta}$ is real as the Hamiltonian must be hermitian, and the condition for the presence of off-diagonal components is similar to that for the Dzyaloshinskii-Moriya interaction between $j$ and $k$ sites~\cite{Moriya60,Dzyaloshinskii58}.

By using the static magnetic susceptibility of electrons between $k$ and $m$ sites, $\chi_{km}^{\alpha\beta}$, we can express the expectation value of the magnetic dipole, $-\gamma_{\rm e}\hbar\bm{S}_{k}$, for the uniform magnetic field $\bm{H}_{m}=\bm{H}$ as
\begin{align}
-\gamma_{\rm e}\hbar S_{k}^{\alpha}=\sum_{m\beta}\chi_{km}^{\alpha\beta}H_{m}^{\beta}
=
\sum_{\beta}\chi_{k}^{\alpha\beta}H^{\beta},
\quad
\chi_{k}^{\alpha\beta}\equiv\sum_{m}\chi_{km}^{\alpha\beta},
\end{align}
where $\gamma_{\rm e}$ is the electron gyromagnetic ratio, and  the local susceptibility tensor $\chi_{k}^{\alpha\beta}$ is symmetric.
Substituting this expression into $H_{\rm hf}^{(j)}$, we obtain
\begin{align}
H_{\rm hf}^{(j)}&=-\frac{1}{\gamma_{\rm e}\hbar}\sum_{k}\sum_{\alpha\beta\gamma}(A_{jk}^{\alpha\gamma}\chi_{k}^{\gamma\beta})I_{j}^{\alpha}H^{\beta}
\cr&
=-\gamma_{\rm n}\hbar\sum_{\alpha\beta}K_{\alpha\beta}^{(j)}I_{j}^{\alpha}H^{\beta},
\end{align}
where we have introduced the shift tensor as
\begin{align}
K_{\alpha\beta}^{(j)}\equiv\frac{1}{\gamma_{\rm n}\gamma_{\rm e}\hbar^{2}}\sum_{k}\sum_{\gamma}A_{jk}^{\alpha\gamma}\chi_{k}^{\gamma\beta}.
\end{align}
$K_{\alpha\beta}^{(j)}$ is in general asymmetric, and the off-diagonal components exist when the vertical mirror symmetry is lost, which is the same condition as the presence of the electric-toroidal dipoles as shown in Table~\ref{tbl:activeq}~\cite{Hayami22}.

Considering $H_{\rm hf}^{(j)}$ and the Zeeman term, the total Hamiltonian $H'$ is given in the form $H'=-\hbar\omega_{0}\sum_{\alpha}I_j^{\alpha}g_{\alpha}^{(j)}$ with $g_{\alpha}^{(j)}=h^{\alpha}+\sum_{\beta}K_{\alpha\beta}^{(j)}h^{\beta}$.
By taking the direction of $\bm{g}^{(j)}$ vector as the quantization axis, the eigenvalues are $\hbar\omega_{m}^{(j)}=-\hbar\omega_{0}|\bm{g}^{(j)}|m$, and the NMR shift is given by
\begin{align}
K^{(j)}=[1+\sum_{\alpha\beta}\{2K_{\alpha\beta}^{(j)}+(\hat{K}^{(j)T}\hat{K}^{(j)})_{\alpha\beta}\}h^{\alpha}h^{\beta}]^{1/2}-1.
\end{align}
In the first-order perturbation, we obtain the first term in (\ref{eq:shiftpb}), and it is clear that the anti-symmetric components contribute from the second order in the perturbation.

\subsection{On expression of nuclear quadrupolar interaction}

The nuclear quadrupolar interaction is given by
\begin{align}
H_{\rm Q}^{(j)}=\frac{eQ}{2I(2I-1)}\sum_{\alpha\beta}V_{\alpha\beta}^{(j)}I_{j}^{\alpha}I_{j}^{\beta},
\end{align}
where $V_{\alpha\beta}^{(j)}=\partial^{2}\phi_{j}/\partial\alpha\partial\beta$.
Since the electrostatic potential at the nuclear site, $\phi_{j}$, satisfies the Laplace equation, $V_{\alpha\beta}^{(j)}$ is traceless, and it is apparent that it is symmetric.
Hereafter, we omit the label $j$ for simplicity.

Using the traceless symmetric property, we have
\begin{align}
\sum_{\alpha\beta}V_{\alpha\beta}I_{\alpha}I_{\beta}=
\sum_{\alpha\beta}V_{\alpha\beta}Q^{\alpha\beta},
\end{align}
where $Q^{\alpha\beta}$ is defined as (\ref{eq:ham}).
Moreover, using ${\rm Tr}\,\hat{V}=0$ and $\bm{I}^{2}=I_{x}^{2}+I_{y}^{2}+I_{z}^{2}$, we have the relation,
\begin{align}
\sum_{\alpha}V_{\alpha\alpha}I_{\alpha}^{2}=
\frac{1}{2}\left[
V_{zz}(3I_{z}^{2}-\bm{I}^{2})
+(V_{xx}-V_{yy})(I_{x}^{2}-I_{y}^{2})
\right].
\end{align}
Introducing the quadrupole operators as
\begin{align}
&
Q_{u}=3I_{z}^{2}-\bm{I}^{2},
\quad
Q_{v}=I_{x}^{2}-I_{y}^{2},
\cr&
Q_{yz}=\frac{1}{2}(I_{y}I_{z}+I_{z}I_{y}),
\quad
Q_{zx}=\frac{1}{2}(I_{z}I_{x}+I_{x}I_{z}),
\cr&
Q_{xy}=\frac{1}{2}(I_{x}I_{y}+I_{y}I_{x}),
\end{align}
and the coefficients,
\begin{align}
V_{u}=\frac{1}{4}V_{zz},
\quad
V_{v}=\frac{1}{4}(V_{xx}-V_{yy}),
\end{align}
we obtain
\begin{align}
&
\sum_{\alpha\beta}V_{\alpha\beta}I_{\alpha}I_{\beta}
=
2\left[
V_{u}Q_{u}
+V_{v}Q_{v}
\right.
\cr&\hspace{1.5cm}\left.
+V_{yz}Q_{yz}
+V_{zx}Q_{zx}
+V_{xy}Q_{xy}
\right].
\label{eq:qi}
\end{align}
Note that in terms of the so-called Stevens' equivalent operators $O_{mn}$~\cite{Stevens52,Huttings64}, we have the correspondences, $Q_{u}=O_{20}$, $Q_{v}=O_{22}$, $Q_{yz}=O_{yz}$, $Q_{zx}=O_{zx}$, $Q_{xy}=O_{xy}$.
In the principal frame where $V_{\alpha\beta}$ is diagonal, it is customary to use the field gradient $q$ and the asymmetry parameter $\eta$, which are defined as $q=V_{zz}/e$, $\eta=V_{v}/V_{u}=(V_{xx}-V_{yy})/V_{zz}$.

Taking the direction of $\bm{h}$ as the quantization axis (denoted by $z'$ axis), we treat the interaction (\ref{eq:qi}) in the first-order perturbation.
Since only the operator $Q_{u}$ in (\ref{eq:qi}) has diagonal elements, $\braket{m|Q_{u}|m}=3m^{2}-I(I+1)$, we obtain the first-order correction as
\begin{align}
\Delta\Omega_{m}=\frac{eQ/\hbar}{4I(2I-1)}V_{z'z'}[3m^{2}-I(I+1)].
\end{align}
Using this expression and $V_{z'z'}=\sum_{\alpha\beta}V_{\alpha\beta}h^{\alpha}h^{\beta}$, we obtain the second term in (\ref{eq:shiftpb}).

\section{On fitting of experimental data}

It is often encountered that the direction of the applied magnetic field with respect to the crystal axes is undetermined in practical experiments.
In this case, we suppose that we rotate the applied magnetic field in the plane defined by the normal vector $\bm{n}=(n_{x},n_{y},n_{z})$ ($|\bm{n}|=1$), which is defined in the orthogonal coordinate $(xyz)$ relating to crystallographic axes $(abc)$.

Let us denote the starting direction of the applied magnetic field as $\bm{s}=(s_{x},s_{y},s_{z})$ ($|\bm{s}|=1$, $\bm{s}\cdot\bm{n}=0$), and the field is rotated by an angle $\vartheta$ in this plane.
Then, the magnetic field direction is expressed as
\begin{align}
\bm{h}&=\hat{C}_{\bm{n}}(\vartheta)\bm{s}
\cr&
=
\begin{pmatrix}
c+n_{x}^{2}\bar{c} & n_{x}n_{y}\bar{c}-n_{z}s & n_{x}n_{z}\bar{c}+n_{y}s \\
n_{x}n_{y}\bar{c}+n_{z}s & c+n_{y}^{2}\bar{c} & n_{y}n_{z}\bar{c}-n_{x}s \\
n_{x}n_{z}\bar{c}-n_{y}s & n_{y}n_{z}\bar{c}+n_{x}s & c+n_{z}^{2}\bar{c}
\end{pmatrix}
\begin{pmatrix} s_{x} \\ s_{y} \\ s_{z} \end{pmatrix},
\cr&
\end{align}
where $\hat{C}_{\bm{n}}(\vartheta)$ is the rotation matrix with respect to $\bm{n}$, and we have used the abbreviations, $c=\cos\vartheta$, $s=\sin\vartheta$, and $\bar{c}=1-c$.
It is easily confirmed that $\bm{n}\cdot\bm{h}=0$ and $\bm{h}=\bm{s}$ for $\vartheta=0$.

Similarly, for the $\varphi$ rotation in the plane perpendicular to $\bm{n}\times\bm{s}$, the magnetic field direction is expressed as
\begin{align}
\bm{h}&=\hat{C}_{\bm{n}\times\bm{s}}(\varphi)\bm{s}.
\end{align}

In this observation, the total shift (\ref{eq:shift}) is regarded as a function of ($\bm{n}$, $\bm{s}$, $k_{i}^{(1)}$, $k_{i}'^{(1)}$, $v_{i}^{(1)}$) in addition to the angles, $\vartheta$ and $\varphi$.
Thus, these variables (fitting parameters) can be determined so as to reproduce the observed total shift $T_{m}^{(j)}(\vartheta)=[\omega_{m-1}^{(j)}(\vartheta)-\omega_{m}^{(j)}(\vartheta)-\omega_{0}]/\omega_{0}$ and optionally $T_{m}^{(j)}(\varphi)$, by using, for example, the multiple nonlinear regression with the backpropagation algorithm in the field of neural networks~\cite{wiki_backpropagation}.
As a result, directions of $\bm{n}$ and $\bm{s}$ are determined in a given coordinate system ($xyz$), in addition to the shift and EFG tensors.


\begin{thebibliography}{99}
\bibitem{Nagaosa23} N. Nagaosa, J. Phys. Soc. Jpn. \textbf{92}, 081002 (2023).
\bibitem{Togawa23} Y. Togawa, A.S. Ovchinnikov, and J. Kishine, J. Phys. Soc. Jpn. \textbf{92}, 081006 (2023).
\bibitem{Yu23} T. Yu, Z. Luo, and G.E.W. Bauer, Phys. Rep. \textbf{1009}, 1 (2023).
\bibitem{Yoda15} T. Yoda, T. Yokoyama, and S. Murakami, Sci. Rep. \textbf{5}, 12024 (2015).
\bibitem{Yoda18} T. Yoda, T. Yokoyama, and S. Murakami, Nano Lett. \textbf{18}, 916 (2018).
\bibitem{Shalygin12} V. A. Shalygin, A. N. Sofronov, L. E. Vorob'ev, and Farbshtein, Phys. Solid State \textbf{54}, 2362 (2012).
\bibitem{Furukawa17} T. Furukawa, Y. Shimokawa, K. Kobayashi, and T. Itou, Nat. Commun. \textbf{8}, 954 (2017).
\bibitem{Furukawa21} T. Furukawa, Y. Watanabe, N. Ogasawara, K. Kobayashi, and T. Itou, \textbf{3}, 023111 (2021).
\bibitem{Oiwa22} R. Oiwa and H. Kusunose, Phys. Rev. Lett. \textbf{129}, 116401 (2022).
\bibitem{Suzuki23} Y. Suzuki and Y. Kato, Phys. Rev. B\textbf{107}, 115305 (2023).
\bibitem{Goehler11} B. G\"ohler, V. Hamelbeck, T. Z. Markus, M. Kettner, G. F. Hanne, Z. Vager, R. Naaman, and H. Zacharias, Science \textbf{331}, 894 (2011).
\bibitem{Naaman12} R. Naaman and D. H. Waldeck, J. Phys. Chem. Lett. \textbf{3}, 2178 (2012).
\bibitem{Naaman19} R. Naaman, Y. Paltiel, and D. H. Waldeck, Nat. Rev. Chem. \textbf{3}, 250 (2019).
\bibitem{Naaman20} R. Naaman, Y. Paltiel, and D. H. Waldeck, J. Phys. Chem. Lett. \textbf{11}, 3660 (2020).
\bibitem{Evers22} F. Evers, A. Aharony, N. Bar-Gill, O. Entin-Wohlman, P. Hedeg\o{a}rd, O. Hod, P. Jelinek, G. Kamieniarz, M. Lemeshko, K. Michaeli, V. Mujica, R. Naaman, Y. Paltiel, S. Refaely-Abramson, O. Tal, J. Thijssen, M. Thoss, J. M. van Ruitenbeek, L. Venkataraman, D. H. Waldeck, B. Yan, and L. Kronik, Adv. Mater. \textbf{34}, 2106629 (2022).
\bibitem{Inui20} A. Inui, R. Aoki, Y. Nishiue, K. Shiota, Y. Kousaka, H. Shishido, D. Hirobe, M. Suda, J.I. Ohe, J.I. Kishine, H. M. Yamamoto, and Y. Togawa, Phys. Rev. Lett. \textbf{124}, 166602 (2020).
\bibitem{Nabei20} Y. Nabei, D. Hirobe, Y. Shimamoto, K. Shiota, A. Inui, Y. Kousaka, Y. Togawa, and H. M. Yamamoto, Appl. Phys. Lett. \textbf{117}, 052408 (2020).
\bibitem{Shiota21} K. Shiota, A. Inui, Y. Hosaka, R. Amano, Y. Onuki, M. Hedo, T. Nakama, D. Hirobe, J.I. Ohe, J.I. Kishine, H. M. Yamamoto, H. Shishido, and Y. Togawa, Phys. Rev. Lett. \textbf{127}, 126602 (2021).
\bibitem{Kelvin04} L. Kelvin, \textit{Baltimore Lectures on Molecular Dynamics and the Wave Theory of Light} (C.J. Clay and Sons, London, 1904).
\bibitem{Wagniere07} G.H. Wagnie\'re, \textit{On Chirality and the Universal Asymmetry: Reflections on Image and Mirror Image} (Wiley-VCH, Weinheim, 2007).
\bibitem{Barron04} L.D. Barron, \textit{Molecular Light Scattering and Optical Activity} (Cambridge University Press, Cambridge, U.K., 2004) 2nd ed.
\bibitem{Kishine22} J. Kishine, H. Kusunose, and H. M. Yamamoto, Isr. J. Chem. \textbf{62}, e202200049 (2022).
\bibitem{Hayami18} S. Hayami, M. Yatsushiro, Y. Yanagi, and H. Kusunose, Phys. Rev. B\textbf{98}, 165110 (2018).
\bibitem{Kusunose20} H. Kusunose, R. Oiwa, S. Hayami, J. Phys. Soc. Jpn. \textbf{89}, 104704 (2020).
\bibitem{Perez22} \'A. Valent\'in-P\'erez, P. Rosa, E.A. Hillard, M. Giorgi, Chirality \textbf{34}, 163 (2022).
\bibitem{Flack83} H.D. Flack, Acta Cryst. A\textbf{39}, 876 (1983).
\bibitem{Flack00} H.D. Flack and G. Bernardinelli, J. Appl. Cryst. \textbf{33}, 1143 (2000).
\bibitem{Flack08} H.D. Flack and G. Bernardinelli, Chirality \textbf{20}, 681 (2008).
\bibitem{Abragam61} A. Abragam, \textit{Principles of Nuclear Magnetism}, (Oxford University Press, Oxford, 1961).
\bibitem{ITA16} M. I. Aroyo, Ed., \textit{International Tables for Crystallography Volume A: Space-group Symmetry}, (Sixth edition, Wiley, 2016).
\bibitem{Koma73} A. Koma, Phys. Stat. Sol. (b) \textbf{56}, 655 (1973).
\bibitem{Koma73a} A. Koma, Phys. Stat. Sol. (b) \textbf{57}, 299 (1973).
\bibitem{Shimizu69} T. Shimizu, Mem. Fac. Techno., Kanazawa Univ. (Japan), \textbf{5}, 189 (1969).
\bibitem{Xue00} X.Y. Xue and M. Kanzaki, Solid State Nucl. Magn. Reson., \textbf{16}, 245 (2000).
\bibitem{Nakamura23} N. Nakamura, A. Yanuma, Y. Chiba, R. Omura, R. Higashinaka, H. Harima, Y. Aoki, and T.D. Matsuda, J. Phys. Soc. Jpn. \textbf{92}, 034701 (2023).
\bibitem{Shimizu20} Y. Shimizu, A. Miyake, A. Maurya, F. Honda, A. Nakamura, Y.J. Sato, D. Li, Y. Homma, M. Yokoyama, Y. Tokunaga, M. Tokunaga, and D. Aoki, Phys. Rev. B\textbf{102}, 134411 (2020).
\bibitem{Miranbet95} F. Mirambet, B. Chevalier, L. Fourn\`es, J.F. Silva, M.A.F. Ramos, and T. Roisnel, J. Mag. Mag. Mat. \textbf{140-144}, 1387 (1995).
\bibitem{Kruk97} R. Kruk, R. Kmie\'c, K. Latka, K. Tomala, R. Tro\'c, and V.H. Tran, Phys. Rev. B\textbf{55}, 5851 (1997).
\bibitem{Ishitobi23} T. Ishitobi and K. Hattori, JPS Annual Metting (2022), 16pGB32-14.
\bibitem{Harima23} H. Harima, SciPost Phys. Proc. \textbf{11}, 6 (2023).
\bibitem{Tabata23} C. Tabata et al., JPS Annual Metting (2022), 16pGB32-13.
\bibitem{Tokunaga23} Y. Tokunaga, Y. Shimizu, H. Sakai, S. Kambe, A. Maurya, F. Honda, A. Nakamura, D.Li, Y. Homma, and D.Aoki, JPS Spring Meeting (2023), 25aH1-12.
\bibitem{Yanagisawa23} K. Tsuchida, R. Hibino, M. Matsuda, H. Hidaka, T. Yanagisawa, H. Amitsuka, C. Tabata, and Y. Shimizu, JPS Annual Meeting (2023), 18aA205-4.
\bibitem{Moriya60} T. Moriya, Phys. Rev. \textbf{120}, 91 (1960).
\bibitem{Dzyaloshinskii58} I. Dzyaloshinskii, J. Phys. Chem. \textbf{4}, 241 (1958).
\bibitem{Hayami22} S. Hayami, R. Oiwa, and H. Kusunose, J. Phys. Soc. Jpn. \textbf{91}, 113702 (2022).
\bibitem{Stevens52} K.W.H. Stevens, Proc. Phys. Soc., Sect. A \textbf{65}, 209 (1952).
\bibitem{Huttings64} M.T. Hutchings, in \textit{Solid State Physics}, ed. F. Seitz and D. Turnbull (Academic Press, New York, 1964) Vol. 16, Chap. 3, p. 227.
\bibitem{wiki_backpropagation} Wikipedia: backpropagation,\\\texttt{https://en.wikipedia.org/wiki/Backpropagation}.
\end{thebibliography}

\end{document}